\documentclass[aps,reprint,,superscriptaddress]{revtex4-2}

\usepackage{graphicx}
\usepackage{bm}
\usepackage{amsmath}
\usepackage{amssymb}
\usepackage{physics}
\usepackage{placeins}
\usepackage{color}
\usepackage[hidelinks,colorlinks=true,linkcolor=blue,urlcolor=blue,citecolor=blue]{hyperref}
\usepackage{hyperref}
\usepackage{cancel}

\usepackage[normalem]{ulem}

\begin{document}

\title{Chiral quantum router with Rydberg atoms}

\author{Nikolaos E. Palaiodimopoulos}
\affiliation{Institute of Electronic Structure and Laser, FORTH, GR-70013 Heraklion, Crete, Greece}

\author{Simon Ohler}
\affiliation{Department of Physics and Research Center OPTIMAS, RPTU Kaiserslautern-Landau, 67663 Kaiserslautern, Germany}

\author{Michael Fleischhauer}
\affiliation{Department of Physics and Research Center OPTIMAS, RPTU Kaiserslautern-Landau, 67663 Kaiserslautern, Germany}

\author{David Petrosyan}
\affiliation{Institute of Electronic Structure and Laser, FORTH, GR-70013 Heraklion, Crete, Greece}

\date{\today}

\begin{abstract}
We exploit controlled breaking of time-reversal symmetry to realize coherent routing of quantum information in spin networks.
The key component of our scheme is a spin triangle whose chirality is determined by the quantum state of a control qubit which thus defines the propagation direction, or a superposition thereof, of the quantum information.
We then consider a particular realization of a coherent router using Rydberg atoms. Our results can facilitate scalable quantum information processing and communication in large arrays of Rydberg atoms.
\end{abstract}

\maketitle

\section{Introduction}
Routing quantum states in spin networks is an important task for efficient quantum information processing and transfer in large quantum processors with finite-range interactions between the qubits. 
Photonic systems are most suitable for transferring quantum states over long distances and several routing schemes have been developed \cite{zhou2013quantum,yan2014single,yuan2015experimental,ahumada2019tunable,lee2022quantum,ren2022nonreciprocal}. 
For short-distance communication tasks, the seminal paper of Bose \cite{bose2003quantum} initiated many studies of coherent transfer of quantum states in networks of coupled spins or qubits \cite{christandl2005perfect,petrosyan2010,nikolopoulos2004electron,burgarth2005conclusive,wojcik2005unmodulated,zwick2014optimized}. In such schemes, routing of quantum information  \cite{zueco2009quantum,chudzicki2010parallel,pemberton2011perfect,paganelli2013routing,zhan2014perfect,christensen2020coherent,qi2021topological} usually involves static or dynamic control of couplings and/or onsite energies of the qubits.
Another possible way to steer the excitation or state transfer is to exploit chiral dynamics that has been associated with breaking the time-reversal symmetry \cite{haldane1988model, wen1989chiral, haldane2008possible}. 
In qubit networks, the first proposal that exploited the breaking of time-reversal symmetry to realize transport in a preselected direction was put forward in \cite{zimboras2013quantum}. Subsequently, a classification of time-symmetric or asymmetric networks was presented in \cite{lu2016chiral}. The breaking of time-reversal symmetry has also been employed to achieve directional state transfer in closed ring geometries \cite{liu2015quantum}, to transfer entanglement on a triangular chain \cite{saglam2023} and to simulate a continuous chiral quantum walk through Floquet engineering \cite{novo2021floquet}. Quantum walk search with time-reversal symmetry breaking have also been studied \cite{wong2015quantum}.

In most of the above schemes, the router is controlled by a classical ``switch'' to direct the quantum information into one of two (or more) possible output channels. But a genuine quantum router (or a transistor) should be controlled by a quantum switch (or a gate) to direct the quantum information
into an arbitrary superposition of propagation channels  
\cite{christensen2020coherent,qspintransistor2016,behera2019designing}.
Here we propose such a genuine quantum router in a spin or qubit network that can be realized with an experimentally relevant system.
The building block of our router is a triangle of interacting spins, or qubits, where the chiral dynamics originates from complex amplitudes of excitation hopping between the spins, controlled by the quantum state of a control qubit. 
We note that the structural and spectral properties of spin triangles with broken time-reversal symmetry have been studied from a graph theoretical perspective \cite{cameron2014} and the emergence of chiral dynamics of excitations in such ``flux'' triangles have been experimentally demonstrated in various setups including nuclear magnetic resonance (NMR) systems \cite{lu2016chiral}, plasmonic nanorings \cite{maleki2022time}, cold atoms \cite{gou2020tunable}, and Rydberg atoms \cite{lienhard2020realization}. 
Inspired by the latter setup that demonstrated state-dependent Peierls phases and chiral motion of excitations, we consider an appropriately configured network of Rydberg atoms and show that this platform can implement a genuine quantum router realizing an arbitrary coherent superpositions of propagation paths for quantum information. 

The paper is organized as follows: 
In Sec. \ref{sec:system} we introduce spin network containing a flux triangle and demonstrate the dynamics of the quantum router.
In Sec. \ref{sec:loc} we describe the envisioned implementation of the system with Rydberg atoms and present the derivation of the effective Hamiltonian verified by exact numerical simulations. Our conclusions are summarized in Sec. \ref{sec:sum}. Details of calculations and additional considerations for the system are deferred to the Appendixes.

\section{Router in a spin network} 
\label{sec:system}

Our quantum router consists of a triangle of spins attached to a longer spin chain. 
We first outline the properties of the spin triangle with complex hopping amplitudes and the conditions to attain chiral dynamics of an excitation. We then discuss how to smoothly launch and absorb a single-excitation wave-packet in a finite spin chain by controlling two spins connected to the opposite ends of the chain. We finally combine the flux-triangle and the spin chain and demonstrate coherent routing of excitation between one sender and two receiver spins. 

\subsection{A flux triangle} 
\label{sec:flux}

\begin{figure}
\includegraphics[width=0.8\columnwidth]{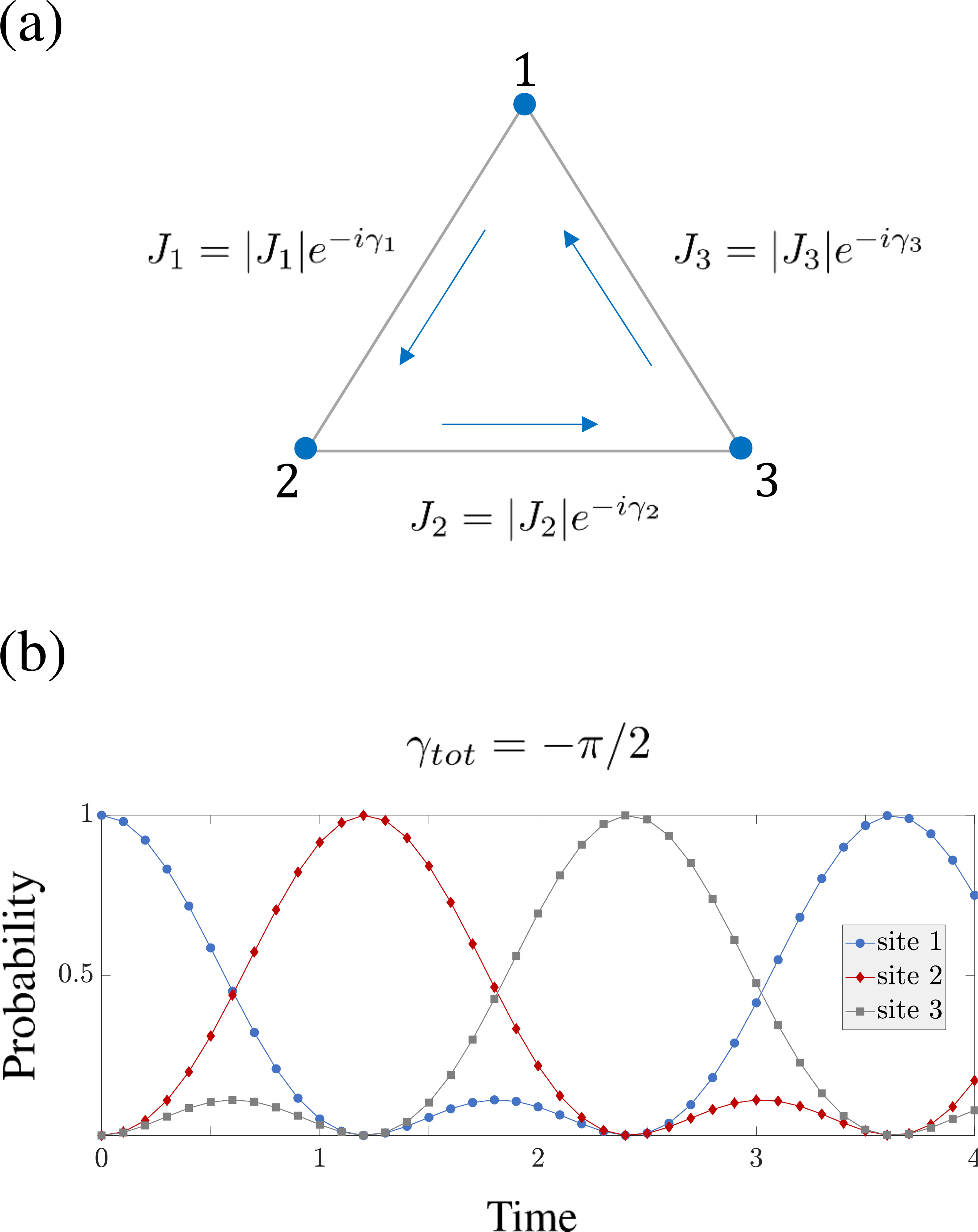}
\caption{(a) Illustration of the spin triangle governed by Hamiltonian (\ref{eq:mat1}). 
(b) Dynamics of populations (excitation  probabilities) of each site of the triangle (time is in units of $1/J$). The excitation is initially localized on site $1$ and the total phase is $\gamma_{\mathrm{tot}}= \gamma_{1}+\gamma_{2}+\gamma_{3}=-\pi/2$ (e.g., $\gamma_{1,2,3}=-\pi/6$) leading to the counter-clockwise circulation of the excitation; for $\gamma_{\mathrm{tot}}=\pi/2$, the circulation direction would be clockwise with the red (site 2) and gray (site 3) curves interchanged.}
\label{fig:ftr}
\end{figure}

The minimal configuration that can act as a quantum state router corresponds to three coupled spins or qubits \cite{lu2016chiral}. Assuming a single excitation, the system is described by the following Hamiltonian matrix in the basis $\{ \ket{100}, \ket{010},\ket{001} \}$ ($\hbar =1$)
\begin{equation} \label{eq:mat1}
H_3= \left(
\begin{array}{ccc}
 0 & J_{1} & J_{3}^* \\
 J_{1}^* & 0 & J_{2} \\
 J_{3} & J_{2}^* & 0\\
\end{array}
\right),
\end{equation}
where the coupling amplitudes $J_{j}$ are complex, $J_{j}=\abs{J_{j}}e^{-i\gamma_{j}}$, see Fig.~\ref{fig:ftr}(a). We may envisage that the phases $\gamma_j$ of the coupling amplitudes stem from an effective magnetic field threading the area of the triangle. 
A detailed analytic treatment of the single-excitation eigenstates and dynamics of the flux triangle can be found, e.g., in Ref. \cite{liu2015quantum}.
Under the conditions 
\begin{subequations} 
\label{eq:conditions}
\begin{eqnarray}
& \abs{J_{1}}=\abs{J_{2}}=\abs{J_{3}}=J , \label{eq:cond1} \\
& \gamma_{\mathrm{tot}}=\gamma_{1}+\gamma_{2}+\gamma_{3}=\pm \pi/2 , \label{eq:cond2}
\end{eqnarray}
\end{subequations}
the dynamics of the system corresponds to sequential localization of the excitation at successive sites, as illustrated in Fig.~\ref{fig:ftr}(b). 
Only the total phase $\gamma_{\mathrm{tot}}$ determines the chiral dynamics of the excitation, while the individual phases $\gamma_j$ can be arbitrary, e.g., we could apply a unitary transformation to make all the phases equal to each other or absorb the total phase into a single coupling element. The sign of the total phase determines the chirality, with $\gamma_{\mathrm{tot}}=+\pi/2$ corresponding to the clockwise and $\gamma_{\mathrm{tot}}=-\pi/2$ to the counter-clockwise circulation of the excitation.
Note that here the switching between output channels is facilitated by the choice of
$\gamma_{\mathrm{tot}}$, i.e., by a prescribed, classical degree of freedom.

\subsection{Non-dispersive wavepacket in a spin chain}
\label{sec:transfer}
We now discuss how to transfer a quantum state via launching and absorbing a quasi-dispersionless wave-packet in a chain of $N$ spins. To this end, we couple in a controllable way a sender $s$ and a receiver $r$ spins to the two ends of the chain. The Hamiltonian of the system is given by ($\hbar=1$)
\begin{equation}
H=H_{0}+H_{sr}(t),
\end{equation}
with
\begin{subequations}
\begin{eqnarray} \label{eq:Hn}
H_{0} &=& \sum_{j=1}^{N-1} J (\hat{\sigma}^{+}_{j} \hat{\sigma}^{-}_{j+1} + \mathrm{H.c.}) + 
\frac{1}{2} \sum_{j=1}^{N} B \hat{\sigma}_{j}^{z}, \\
\label{eq:Ht}
H_{sr}(t) &=& [J_{s}(t)\hat{\sigma}^{+}_{1} \hat{\sigma}^{-}_{s}+ J_{r}(t)\hat{\sigma}^{+}_{N} 
\hat{\sigma}^{-}_{r}+ \mathrm {H.c.}] 
\nonumber \\ 
& & +\frac{1}{2}(B_{s} \hat{\sigma}_{s}^{z}+B_{r} \hat{\sigma}_{r}^{z}) ,
\end{eqnarray}
\end{subequations}
where $\hat{\sigma}^{x,y,z}_{j}$ are the Pauli spin operators, $\hat{\sigma}^{\pm}_j = \frac{1}{2} (\hat{\sigma}^x_j \pm i \hat{\sigma}^y_j)$ are the raising and lowering operators, $J$ is the uniform coupling between the nearest-neighbor spins of the chain, and $B$ is the effective magnetic field which is assumed uniform and can be set to zero without loss of generality (meaning we work in the frame rotating with frequency $B$). 
$H_{sr}(t)$ describes the time-dependent couplings of the sender and receiver spins, $J_{s,r}(t)$, to the first and last spins of the chain, and the effective magnetic fields $B_{s,r}$ for tuning their transition frequencies.

Initially, at $t=0$, a qubit state is encoded in the sender spin, $\ket{\psi} = c_0 \ket{\uparrow}_s + c_1 \ket{\downarrow}_s$, while all the spins of the intermediate chain are prepared in state $\ket{\downarrow}_i$, and our aim is to retrieve the state from the receiver spin at time $t=T$. Since the total spin $\hat{\Sigma}_{z}=\sum_{j=1}^{N} \hat{\sigma}^{z}_{j}$ commutes with the Hamiltonian  $[H,\hat{\Sigma}_{z}]=0$, the complete Hilbert space of the system can be decomposed into decoupled subspaces, each having a fixed number of spin-up excitations. Following the common approach \cite{bose2003quantum,christandl2005perfect} for transferring a quantum state between two spins $s$ and $r$, we consider only the single-excitation subspace and calculate the efficiency of the excitation transfer via
\begin{equation}
P_{T}=\abs{\bra{r}e^{-iHT}\ket{s}}^2,
\end{equation}
which gives the probability of an excitation initially localized at the sender $\ket{s} \equiv \ket{\uparrow_s, \downarrow_1,\ldots,  \downarrow_N, \downarrow_r}$ to be retrieved from the receiver $\ket{r} \equiv \ket{\downarrow_s, \downarrow_1,\ldots,  \downarrow_N, \uparrow_r}$ at time $T$. In order to achieve coherent transfer of any superposition state -- not just the excitation -- it is also important that the phase $\zeta=\mathrm{arg} \big( \bra{r}e^{-iHT}\ket{s} \big)$ acquired during the evolution be fixed and known. 

The single-excitation energy spectrum of the intermediate chain with nearest-neighbor exchange \cite{bose2003quantum,christandl2005perfect,nikolopoulos2004electron} is   
\begin{equation} \label{eq:cos}
    E_{n}=2J\cos(\frac{\pi n}{N+1}), \quad n=1,2...,N . 
\end{equation}
For a chain with longer-range exchange interactions that we will encounter below, this spectrum is slightly modified but can be treated similarly (see Appendix \ref{sec:LSLR} and Eq.~(\ref{eq:sumcos}) there).
To launch a quasi-dispersionless wave-packet, the sender spin should couple predominantly to the linear part of the spectrum, i.e., around $E_{n}=0$, which is analogous to a requirement of the classical wave theory that the second derivative of the dispersion relation (group velocity dispersion) be zero for a wave-packet to maintain its shape during propagation. We therefore tune $B_{s}=0$ and adjust the coupling strength $J_{s}$ to avoid leakage to eigenstates that do not reside near the linear part of the spectrum. It has been shown \cite{osborne2004propagation,banchi2010optimal} that, for a static coupling between the sender and the chain, there is an optimal coupling strength, $J_{s}=J N^{-1/6}$ that ensures that the generated wavepacket does not significantly broaden during propagation through the chain. 

Here, instead, we consider time-dependent coupling $J_{s}(t)$ having linear temporal profile,
\begin{equation} \label{eq:times}
J_{s}(t)= \begin{cases}
J\frac{t}{T} & 0 \leq t \leq t_{m}\\
0 & t_{m} < t\leq T
\end{cases}
,
\end{equation}
where $t_{m}$ is the duration of modulation. By adjusting $t_{m}$, we can ``smoothly" couple the excitation to a few eigenstates in the linear part of the spectrum, creating therefore a highly-localized wavepacket in momentum space (around $|k|=\pi/2$) but broad (long) in real space (time). We then let the system evolve for time $T$, while applying at time $t_m'=T-t_{m}$ the time-reversed modulation to the coupling of the receiver spin, 
\begin{equation} \label{eq:timer}
J_{r}(t)= \begin{cases}
0 & 0 \leq t < t_m'\\
J(1-\frac{t}{T})  & t_m' \leq t \leq T
\end{cases}
.
\end{equation}
The receiver spin thus fully absorbs the incoming wavepacket, provided it is also resonant with the linear part of the spectrum, $B_{r}=0$.

As for the phase accumulated during the transfer, if we start with an excitation localized at the sender site and let the system evolve, the amplitude of the wavefunction $\propto e^{ikj}$ will acquire a $k=-\pi/2$ phase factor for every step along the chain \cite{petrosyan2010,pemberton2010quantum}.
The total phase at the receiver site will then be $\zeta=(-\pi/2)(N+1) \mod 2\pi$.

\subsection{The complete setup}
\label{sec:modelsetup}

\begin{figure*}[ht]
\includegraphics[width=0.8\textwidth]{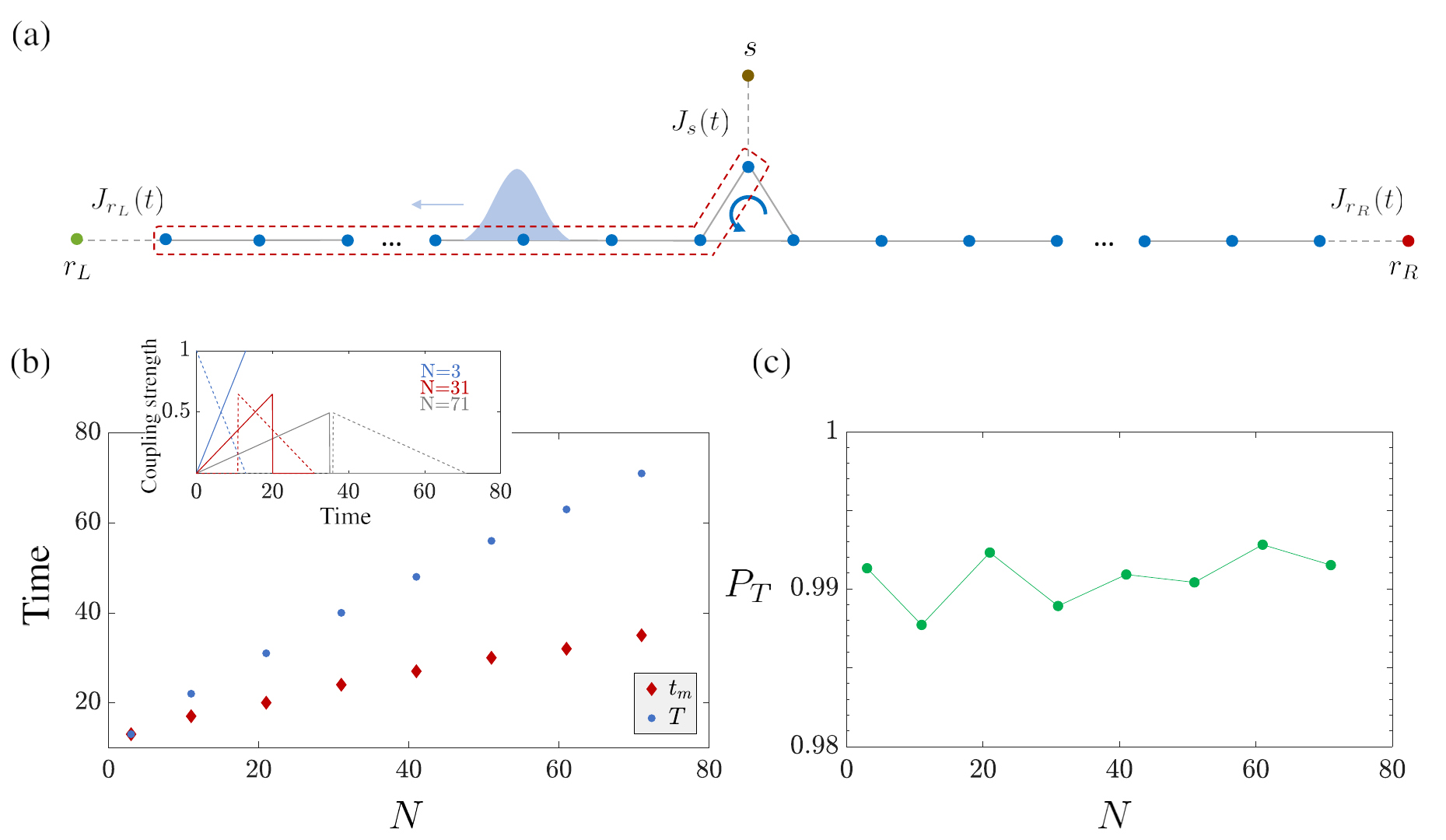}
\caption{(a) Schematics of the proposed setup for quantum router in a network of spins involving the flux triangle with $\gamma_{\mathrm{tot}}=-\pi/2$ and spin chain coupled to the sender and receiver spins. 
(b) Modulation time $t_m$ and the total transfer time $T$ vs the length $N$ of the spin chain (including two sites of the flux triangle) enclosed by the red dashed box in (a). Inset illustrates the temporal profiles of the coupling strengths $J_{s}$ (solid lines) and $J_{r_L,r_R}$ (dashed lines) for three different $N=3,31,71$. 
(c)~Transfer probability $P_{T}$ at $t=T$ for different lengths $N$ of the chain. Time is in units of $1/J$ and coupling strength $J_{s,r_L,r_R}$ are in units of $J$.}
\label{fig:rout1}
\end{figure*}

Our router consists of the flux-triangle whose two sites are part of a chain of spins, as shown in Fig. \ref{fig:rout1}(a). We connect the sender spin $s$ to the apex of the triangle with coupling amplitude $J_{s}$ and two receiver spins $r_{L,R}$ to the opposite ends of chain with coupling amplitudes $J_{r_L,r_R}$. 
Initially the sender spin is excited and all the other spins are in their ground state, and our goal is to efficiently transfer the excitation to one of the receiver spins. To this end, we modulate the coupling $J_{s}(t)$ in order to generate a wavepacket that will travel to the left or to the right part of the spin chain depending on the total phase $\gamma_{\mathrm{tot}}$ of the flux triangle. As the wavepacket propagates in the chain, we modulate the couplings $J_{r_L,r_R}(t)$ of the receiver spins to absorb the incoming wavepacket. 

The direction of propagation of the wavepacket launched by the sender spin is determined by the total phase $\gamma_{\mathrm{tot}}$ in the flux triangle \cite{zimboras2013quantum}. For $\gamma_{\mathrm{tot}} = -\pi/2$ the wavepacket is sent to the left of the chain, while for $\gamma_{\mathrm{tot}} = \pi/2$ it is sent to the right. This can be intuitively understood by considering the phase that the excitation acquires at each site upon the evolution. As stated above, for $J \in \mathbb{R}^{+}$ at each hopping step the amplitude of the wavefunction acquires a factor of $e^{-i\pi/2}$. But if the couplings $J_i$ between the sites of the triangle are complex, their arguments must also be accounted for when calculating the acquired phase along a particular path. In the flux-triangle, there are two paths for an excitation initially localized on site 1 to reach site 2: either direct hopping $1 \to 2$, or in two steps $1 \to 3 \to 2$ via site 3. The acquired phase factors for the two paths are $e^{-i(\pi/2-\gamma_{1})}$ and $e^{-i(\pi+\gamma_{3}+\gamma_{2})}$ respectively. Analogously, we can calculate the acquired phase factors for the two paths leading from site 1 to site 3. If the conditions (\ref{eq:conditions}) are satisfied with $\gamma_{\mathrm{tot}}=-\pi/2$, we have constructive interference on site $2$ and destructive interference on site $3$, while the opposite holds true for $\gamma_{\mathrm{tot}}=\pi/2$. In Fig. \ref{fig:ftr}(b) we observe that, for $\gamma_{\mathrm{tot}}=-\pi/2$, site 3 is weakly populated before the wavefunction is localized on site 2, which is due to the fact that it takes longer to reach site 3 via path $1 \to 2 \to 3$ and interfere destructively. 

Hence, the transfer scheme presented in Sec. \ref{sec:transfer} can be applied to the setup of Fig. \ref{fig:rout1}(a).
By gradually turning on the coupling of the sender spin to the apex of the flux triangle, we avoid significantly populating site $3$ before the destructive interference takes place. The launched wavepacket, provided it is narrow in momentum space around $k=-\pi/2$ (sufficiently smooth and long in time), is then sent almost solely to the left of the chain
(see Fig. \ref{fig:rout1}(a)). For $\gamma_{\mathrm{tot}}=\pi/2$, the launched wavepacket would propagate to the right.
To ensure that the sender spin is resonantly coupled to the linear part of the spectrum of the chain (left or right part), for convenience we consider configurations with odd number of spins in each subchain (i.e. a zero-energy eigenstate is always present). In Fig.~\ref{fig:rout1}(b) we show the total transfer time $T$ and the modulation time $t_{m}$ versus the length of the intermediate chain while in the inset we illustrate the temporal profiles of the coupling strengths $J_{s}(t)$, $J_{r_L}(t)$ for some indicative chain lengths. 
Note that for short chains we have $t_{m} \approx T$ and the temporal modulation of the boundary couplings resembles the counter-intuitive pulse sequence of the stimulated Raman adiabatic passage (STIRAP) scheme \cite{STIRAP-RMP1998,vitanov2017stimulated}, where the pulse-overlap is optimized to ensure adiabaticity. In our transfer scheme, however, once the chain is sufficiently long, the couplings $J_{s}(t)$, $J_{r_L,r_R}(t)$ need not have temporal overlap. In Fig.~\ref{fig:rout1}(c) we show the corresponding transfer probabilities for different lengths of the intermediate chain, attesting to the excellent scalability and efficiency of our transfer scheme. 
The acquired phases during the transfer are $\zeta_{L}=-(N+1)(\pi/2)+\gamma_{1}$ and $\zeta_{R}=-(N+1)(\pi/2)-\gamma_{3}$ for the left or the right receiver spin with $\gamma_\mathrm{tot} = \mp \pi/2$, respectively.

So far, we have described a router that can transfer a quantum state of the sender spin to the distant left or right receiver spin, determined by the total phase of the flux-triangle. But to realize a genuine quantum router, the directionality of the transfer should be determined by the state of a control qubit, which in general can be in a superposition state and therefore create distributed entanglement in the system. We will address this issue in the next section.

\section{Implementation of a quantum router with Rydberg atoms} 
\label{sec:loc}

\begin{figure*}
\includegraphics[width=1.0\textwidth]{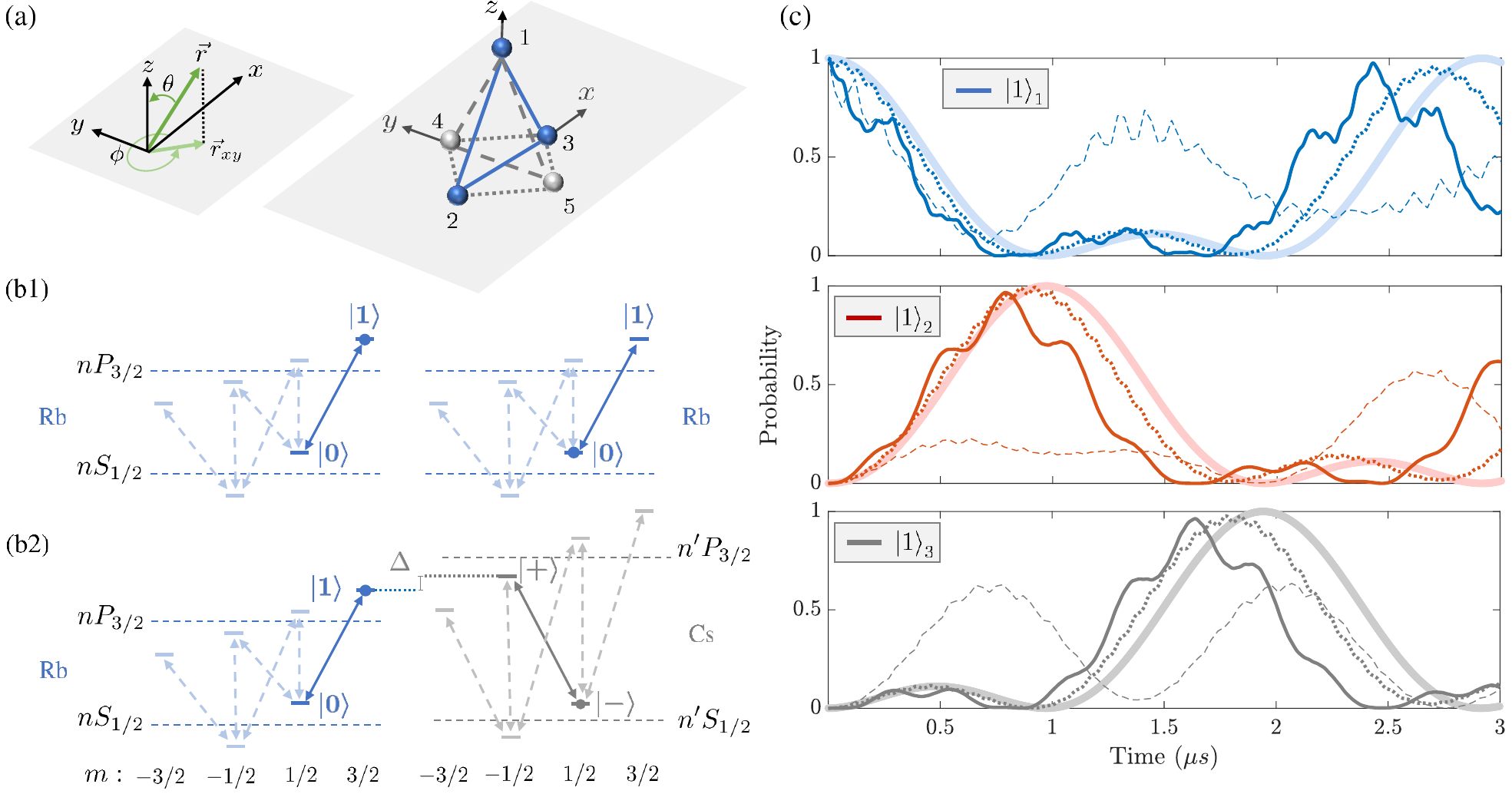}
\caption{(a) Spatial arrangement of the main (1,2,3, blue) and auxiliary (4,5, gray) atoms to realize the flux tiangle. 
(b) Rydberg state manifold for two Rb atoms (b1) and Rb and Cs atoms (b2) in a static magnetic field that lifts the degeneracy of the magnetic sublevels $m$. 
(c) Dynamics of populations of the excited states $\ket{1}_{i}$ of Rb atoms $i=1,2,3$ of the triangle sharing a single excitation, as obtained from the effective Hamiltonian (\ref{eq:mat3}) (thick faded lines), for the truncated system of four two-level atoms (dotted lines), and for the complete system of four six-level atoms without optimization (dashed lines).
We choose the principal quantum numbers $n=70$ for Rb and $n'=71$ for Cs, set the distance $r_{23} = a=17\, \mu$m and solve Eqs.~(\ref{eq:sys}) obtaining $r_{24}=r_{34}=r_{52}=r_{53}=b = 9.39\, \mu$m ($r_{45}=4 \, \mu$m) and $r_{12}=r_{13}=c=10.04\, \mu$m, while $\Delta= 2\pi \times 14.29\,$MHz for $B=26.84 \, \mathrm{G}$, leading to $\abs{J_{12}}=\abs{J_{23}}=\abs{J_{13}}=2\pi \times 198.1 \,$kHz with the corresponding angles $\gamma_{12}=\gamma_{13}=-0.151\pi$, $\gamma_{23}=-0.198\pi$. 
We then optimize the complete system of four six-level atoms to attain good chiral dynamics (solid lines).
The parameters obtained by the optimization are $r_{23} = a=17\, \mu$m, $r_{24}=r_{34}=r_{52}=r_{53}=b = 12.25\, \mu$m ($r_{45}=8.8 \, \mu$m) and $r_{12}=r_{13}=c=9.83\, \mu$m, while $\Delta= 2\pi \times 2.68\,$MHz for $B=46.38 \, \mathrm{G}$, leading to $\abs{J_{12}}=\abs{J_{13}}=2\pi \times 238 \,$kHz and $\abs{J_{23}}=2\pi \times 265.7 \,$kHz with corresponding angles $\gamma_{12}=\gamma_{13}=-0.2548\pi$, $\gamma_{23}=0.0113\pi$.}
\label{fig:ryd}
\end{figure*}

We envision a realization of the flux triangle and the spin network with Rydberg atoms.
By making use of the density-dependence of complex hopping amplitudes in 
Rydberg systems \cite{lienhard2020realization,ohler2022self}, we can implement a genuine quantum router, i.e., a switch that directs the transport of quantum information depending on the quantum state of a control qubit.

Consider the geometry shown in Fig.~\ref{fig:ryd}(a), where we assume two different kinds of atoms (blue and gray). 
The three spins of the triangle correspond to atoms 1,2,3 (blue) in the $xz$-plane, and we place two auxiliary atoms 4,5 (gray) in the $xy$-plane. One or the other auxiliary atom will mediate second-order excitation transfer between the main atoms, similar to \cite{lienhard2020realization}, leading to complex exchange interactions, as detailed below. 
To be specific, we assume that the main atoms of the triangle and the rest of the spin chain are Rb (blue), while the auxiliary atoms are Cs (gray). Other choices are also possible, but having different atomic species can facilitate in the experiments their selective laser excitation to the Rydberg states. 

\subsection{Dipole-dipole interaction} \label{sec:rsm}

A pair of atoms $i$ and $j$ in Rydberg states interact with each other via the dipole-dipole interaction
\begin{equation} \label{eq:dipole}
V_{ij}=\frac{1}{4\pi \epsilon_{0} |r_{ij}|^{3}}\big[\hat{\bm{d}}_{i} \cdot \hat{\bm{d}}_{j}-3( \hat{\bm{d}}_{i} \cdot \hat{\bm{r}}_{ij}) (\hat{\bm{d}}_{j} \cdot \hat{\bm{r}}_{ij})\big],
\end{equation}
where $\hat{\bm{d}}_{i}= \hat{e}_x \hat{d}_{i}^{x} + \hat{e}_y \hat{d}_{i}^{y} + \hat{e}_z \hat{d}_{i}^{z} = \hat{e}_+ \hat{d}_{i}^{+} + \hat{e}_- \hat{d}_{i}^{-} + \hat{e}_z \hat{d}_{i}^{z}$ is the (vector) dipole operator with $\hat{e}_{\pm} = \mp(\hat{e}_x\pm i \hat{e}_y)/\sqrt{2}$, $\hat{\bm{r}}_{ij} = \vec{r}_{ij}/\abs{\vec{r}_{ij}}$ is the unit vector along the relative position vector $\vec{r}_{ij}$ and $r_{ij}=\abs{\vec{r}_{ij}}$ is the interatomic distance. In Fig.~\ref{fig:ryd}(b1-b2) we show the Rydberg state manifold with the relevant atomic transitions between the levels $nS_{1/2}$, $nP_{3/2}$ of Rb and $n^{\prime}S_{1/2}$, $n^{\prime}P_{3/2}$ of Cs. The unperturbed energy levels of an atom in Rydberg state with the principal quantum number $n$ are 
\begin{equation}
E_{nl}=-\frac{\mathrm{Ry}}{(n-\delta_{l})^2} , \label{eq:RydbE}   
\end{equation}
where $\mathrm{Ry}$ is the Rydberg constant and $\delta_{l}$ is the quantum defect that depends on the atomic species and the orbital angular momentum $l$ of the Rydber state electron.
We assume that a sufficiently strong static and homogeneous magnetic field $B$ along the $z$ direction defines the quantization axis and lifts the degeneracy of the magnetic sublevels. 
For simplicity, we consider linear Zeeman splitting of the magnetic sublevels $\delta E_{jm}=\mu_{B} B g_j m$, where $\mu_{B}$ is the Bohr magneton, $g_j$ is the Lande factor and $m$ is the projection of the total angular momentum $j$ onto the quantization axis. 
 
Let us first assume that each atom has only one relevant transition between a pair of levels: $\ket{0} \to \ket{1}$ ($\Delta m=+1$) for Rb, and $\ket{-} \to \ket{+}$ ($\Delta m=-1$) for Cs, as illustrated in Fig.~\ref{fig:ryd}(b1-b2) with solid lines; later on we will account for all the levels and transitions of the atoms.
The pairwise dipole-dipole interactions between the main atoms $V^{(\mathrm{AA})}$ and between the main and auxiliary atoms $V^{(\mathrm{AB})}$ then reduce to (see Appendix \ref{sec:appDDI})
\begin{subequations}\label{eq:dd12}
\begin{equation}\label{eq:dd1}
\begin{split}
V^{(\mathrm{AA})}_{ij}= & \frac{|d_\mathrm{A}^+|^2}{4\pi \epsilon_{0} |r_{ij}|^{3}} 
\frac{1}{2} \big( \ket{0}_i \bra{1} \otimes  \ket{1}_j \bra{0}\\ 
&+ \ket{1}_i \bra{0} \otimes  \ket{0}_j \bra{1} \big) \times  (1-3\cos^{2}{\theta_{ij}}),
\end{split}
\end{equation}
\begin{equation}\label{eq:dd2}
\begin{split}
V^{(\mathrm{AB})}_{ij} &= - \frac{|d_\mathrm{A}^+ d_\mathrm{B}^+|}{4\pi \epsilon_{0} |r_{ij}|^{3}} 
\frac{3}{2} \big( \ket{1}_i \bra{0} \otimes \ket{-}_j \bra{+} e^{-2i\phi_{ij}}\\ 
& \qquad +  
\ket{0}_i \bra{1} \otimes \ket{+}_j \bra{-} e^{2i\phi_{ij}} \big) \sin^{2}{\theta_{ij}} ,
\end{split}
\end{equation}  
\end{subequations}
where $\theta_{ij}$ is the angle between the relative position vector $\vec{r}_{ij}$ and the quantization axis $\hat{z}$, 
$\phi_{ij}$ is the polar angle (see Fig.~\ref{fig:ryd}(a)), and $d_{\mathrm{A,B}}^{\pm}$ are the dipole matrix elements of the corresponding transitions of the main (A) and auxiliary (B) atoms. 
Note that in the above equations the $\theta$-dependence is relevant only for atom $1$, since the rest of the atoms are in the $xy$-plane and therefore $\theta=\pi/2$; while the $\phi$-dependence is relevant only for the interaction between the main and auxiliary atoms involving $\Delta m^{(\mathrm{A,B})} = +1$ or $-1$ transitions (see Appendix \ref{sec:appDDI}). 

We assume that initially one main atom of the triangle is in the excited state $\ket{1}$ and the rest of the atoms are in their lower states $\ket{0}$ and $\ket{-}$.
The transition $\ket{-}\to \ket{+}$ of the auxiliary atoms is strongly detuned by $\Delta \gg  | \langle V^{(\mathrm{AB})} \rangle |$ with respect to the transition $\ket{0} \to \ket{1}$ of the main atoms.  
We also assume that only one of the auxiliary atoms, 4 or 5, is prepared initially in the Rydberg state $\ket{-}$ by the control qubit (see below and Appendix \ref{sec:RBGate}), while the other atom is in its ground electronic state and does not participate in the interaction.

\subsection{Effective model for the flux triangle} 
\label{sec:ed}

The excitation hopping between any pair of the main atoms $i$ and $j$ of the triangle involves two processes: 
(\textit{i}) Resonant excitation exchange $\ket{1}_i\ket{0}_j \to \ket{0}_i\ket{1}_j$ between the main atoms described by Eq. (\ref{eq:dd1}) and illustrated Fig.~\ref{fig:ryd}(b1); and
(\textit{ii}) excitation exchange between an excited main atom and an auxiliary atom described by Eq.~(\ref{eq:dd2}) and shown in Fig.~\ref{fig:ryd}(b2) followed by an exchange between the auxiliary atom and another main atom, $\ket{1}_i \ket{-}_k \ket{0}_j \to \ket{0}_i \ket{+}_k \ket{0}_j \to \ket{0}_i \ket{-}_k \ket{1}_j$.
Since we assume large detuning $\Delta$, the non-resonant state $\ket{+}_k$ of the auxiliary atom $k$ is only virtually populated and can be eliminated adiabatically, leading to the second-order exchange interaction between the main atoms $i$ and $j$
which is responsible for the complex phase of the transition amplitudes determined by the polar angles $\phi$ in Eq.~(\ref{eq:dd2}). 
We thus obtain an effective Hamiltonian for the three main atoms 1,2,3 in the single-excitation basis $\{ \ket{100}, \ket{010},\ket{001} \}$ as
\begin{equation} \label{eq:mat3}
H_{\mathrm{eff}}=
\left(
\begin{array}{ccc}
\mu_{1} & \abs{J_{12}} e^{-i \gamma_{12}} & \abs{J_{13}} e^{i\gamma_{13}}\\
 \abs{J_{12}} e^{i \gamma_{12}} & \mu_{2} & \abs{J_{23}} e^{-i \gamma_{23}}\\
 \abs{J_{13}} e^{-i \gamma_{13}} & \abs{J_{23}} e^{i \gamma_{23}} &  \mu_{3}
\end{array}
\right),
\end{equation}
with
\[
\begin{split}
&\mu_{i}= S^{(2)}_{i \to k \to i} , \\ 
&\abs{J_{ij}}=\abs{T_{i\to j} + T^{(2)}_{i \to k \to j}} , \\
& \gamma_{ij}=\arg \left(T_{i\to j} + T^{(2)}_{i \to k \to j} \right) ,
\end{split}
\]
where 
\[ 
T_{i\to j} = \frac{C_\mathrm{AA}}{r_{ij}^3} \frac{1}{2} (1-3\cos^2{\theta_{ij}}), \quad  
C_{\mathrm{AA}} \equiv \frac{|d_\mathrm{A}^+|^2}{4\pi \epsilon_0}
\]
are the amplitudes of the resonant dipole-dipole exchange interaction between the main atoms $i$ and $j$, 
while
\begin{eqnarray*}
S^{(2)}_{i \to k \to i} &=& -\frac{1}{\Delta} \frac{C_{\mathrm{AB}}^2}{r_{ik}^6} \left(\frac{3}{2}\right)^2 \sin^{4}{\theta_{ki}} , 
\quad C_{\mathrm{AB}} \equiv \frac{|d_\mathrm{A}^+ d_\mathrm{B}^+|}{4\pi \epsilon_{0}} , \\
T^{(2)}_{i \to k \to j} &=& -\frac{1}{\Delta} \frac{C_{\mathrm{AB}}^2}{r_{ik}^3 r_{kj}^3 } \left(\frac{3}{2}\right)^2 e^{-2 i \phi_{ik}} e^{2 i \phi_{kj}} \sin^{2}{\theta_{ik}} \sin^{2}{\theta_{kj}}
\end{eqnarray*}
are the second-order level shifts of, and excitation exchange between, the main atoms via virtual excitation of the auxiliary atom $k=4$ or $5$.

To implement the flux triangle, we look for the parameters of the system that would satisfy conditions (\ref{eq:conditions})
while $\mu_{1}=\mu_{2}=\mu_{3}$. To this end, we choose the principal quantum numbers $n$ and $n'$ for Rb and Cs atoms leading to a small difference ($\sim 2\pi \times 30$ MHz) in their $nS_{1/2} \to nP_{3/2}$ transition frequencies  as per Eq.~(\ref{eq:RydbE}), and calculate their dipole moments $d^{+}_{\mathrm{A}}$ and $d^{-}_{\mathrm{B}}$ on the transitions $\ket{0} \to \ket{1}$ and $\ket{-} \to \ket{+}$ (see Appendix \ref{sec:appDDI}).
Next, we fix the distance $r_{23}=a$ between atoms $2$ and $3$, and place the auxiliary atoms 4 and 5 at equidistant positions from atoms $2$ and $3$, $r_{24}=r_{34}=r_{52}=r_{53}=b$, so that $\mu_{2}=\mu_{3}$.
We place atom $1$ at an equidistant position from atoms $2$ and $3$, $r_{12}=r_{13}=c$ and note that $c \neq b$.  
Then, assuming only the presence of the auxiliary atom 4, we solve numerically the system of equations 
\begin{equation}
\label{eq:sys}
\begin{split}
    & \mu_{1}=\mu_{2,3} , \\
    & \abs{J_{12,13}}=\abs{J_{23}} , \\
    &  \gamma_{\mathrm{tot}} = \gamma_{12} + \gamma_{23} +\gamma_{13}=-\pi/2 , 
\end{split}    
\end{equation}
obtaining thereby the distances $b,c$ (for a fixed $a$) and the detuning $\Delta$ which determines the applied magnetic field $B$. 

In Fig.~\ref{fig:ryd}(c) we show the results for one such solution, where we compare the dynamics of the excitation in the triangle, as obtained for the effective three-state system of Eq.~(\ref{eq:mat3}) (thick faded lines) and for the four level system (dotted lines) including the transition $\ket{-} \to \ket{+}$ of the auxiliary atom $4$. We observe good agreement between the effective and exact (but truncated) models with nearly perfect chiral dynamics of the excitation circulating in the counter-clockwise direction. 
The exact (but truncated) model exhibits slightly different circulation frequency, due to small violation of the conditions for adiabatic elimination of state $\ket{+}_4$, and small-amplitude large-frequency oscillations associated with the finite detuning $\Delta$ of level $\ket{+}_4$. 
But when we include in our calculations all the levels and allowed transitions of the four six-level atoms sharing a single excitation (Hilbert space size is $128$), we obtain complete distortion of the desired chiral dynamics, see Fig.~\ref{fig:ryd}(c) (dashed lines), which is due to the additional level shifts and higher-order transitions of the atoms induced by various non-resonant transitions in the system. 

There are two possibilities to circumvent this problem.
One option is to choose lower principal quantum numbers $n,n^{\prime}$ or/and to increase the distances between the atoms so that the dipole-dipole exchange interactions are much smaller than the detunings of all the spurious non-resonant states which therefore do not distort much the dynamics of the effective three-level or truncated four-level system. But then the effective couplings $\abs{J_{ij}}$ between the triangle atoms will be weak and the resulting slow dynamics will suffer much from the decay of the Rydberg states. 

A better option is to start with the exact solution for the effective model, include all the relevant atomic levels and transitions and then optimize the distances $b,c$ and the magnetic field $B$ so that the chiral dynamics still persists in the full system (four six-level atoms). In Fig.~\ref{fig:ryd}(c) we show one such solution (solid lines) obtained by an optimization algorithm that uses the Nelder-Mead method \cite{NelderMead}.
Even though this solution deviates significantly from the smooth analytic dynamics of the effective system, it still exhibits chiral dynamics on a fast time scale compared to the lifetimes of the Rydberg states. 

We note that if, instead of the auxiliary atom $4$, we place atom $5$, due to the symmetry of the setup, the excitation in the triangle circulates in the clockwise direction.

\subsection{The complete setup}

Following the blueprint of Sec.~\ref{sec:modelsetup}, we now describe the complete setup. We assume that atoms in an array of microtaps are selectively excited by the lasers to the Rydberg state $\ket{0}$ to form the required network for manipulation, routing and transfer of quantum states between the sender and receiver spins -- qubits. 

To set the phase $\gamma_{\mathrm{tot}} = \pm \pi/2$ of the flux triangle, we need to excite one of the auxiliary atoms, 4 or 5, to state $\ket{-}$, or create an arbitrary superposition of the form $\alpha \ket{-}_5 + \beta \ket{-}_4$ that would lead to the corresponding superposition of the propagation directions of the excitation wavepacket, or quantum state of the sender spin, to the right or left receiver spin. We assume that the control qubit, that determines which of the two auxiliary atoms 4 or 5 is prepared in the Rydberg state $\ket{-}$, is encoded in a ground state superposition of one of the auxiliary atoms, e.g., atom 4, $\ket{\psi}_4 = \alpha \ket{g}_4 + \beta \ket{e}_4$, while the other auxiliary atom is in state $\ket{e}_5$.  
Then, by using the sequence of laser pulses as in the standard Rydberg blockade gate \cite{Lukin2001,SWMRMP2010}, we can prepare the corresponding superposition of single Rydberg excitation of the two auxiliary atoms, provided they can be individually addressed by the lasers while strongly interacting when excited to the Rydberg state (see Appendix \ref{sec:RBGate}).

The dynamical control of the couplings $J_{s,r}(t)$ of the sender and receiver spins to the spin network, as per Eqs.~(\ref{eq:times},\ref{eq:timer}), can be achieved in several ways. Conceptually the simplest, but experimentally involved, method would be to move, during time $0\leq t  < t_m$, the sender atom, encoding a qubit in the Rydberg state superposition $c_0 \ket{0}_s + c_1\ket{1}_s$, from some large distance to the position next to the first atom of the network; and move, during time $t_m' \leq t \leq T$, the receiver atom, initially in the Rydberg state $\ket{0}_r$, away from the last atom of the network. 
A more practical approach for static atoms trapped at appropriate positions is to encode the qubit in the ground state superposition $c_0 \ket{g}_s + c_1\ket{e}_s$ of the sender atom, apply a resonant $\pi$-pulse on the transition $\ket{g}_s \to \ket{0}_s$ and then apply a (near-)resonant laser pulse $\Omega_s(t)$ on the transition  $\ket{e}_s \to \ket{1}_s$ to the excited Rydberg state, from where the excitation will hop in the spin network of the Rydberg atoms, until it reaches the receiver atom initially in state $\ket{0}_r$, which is then transferred to the qubit state $\ket{e}_r$ by another laser pulse $\Omega_r(t)$ acting on the transition  $\ket{1}_r \to \ket{e}_r$ followed by resonant $\pi$-pulses on the transitions $\ket{0}_{s,r} \to \ket{g}_{s,r}$ (see Appendix \ref{sec:Jtsr}).  

\begin{figure}
\includegraphics[width=1\columnwidth]{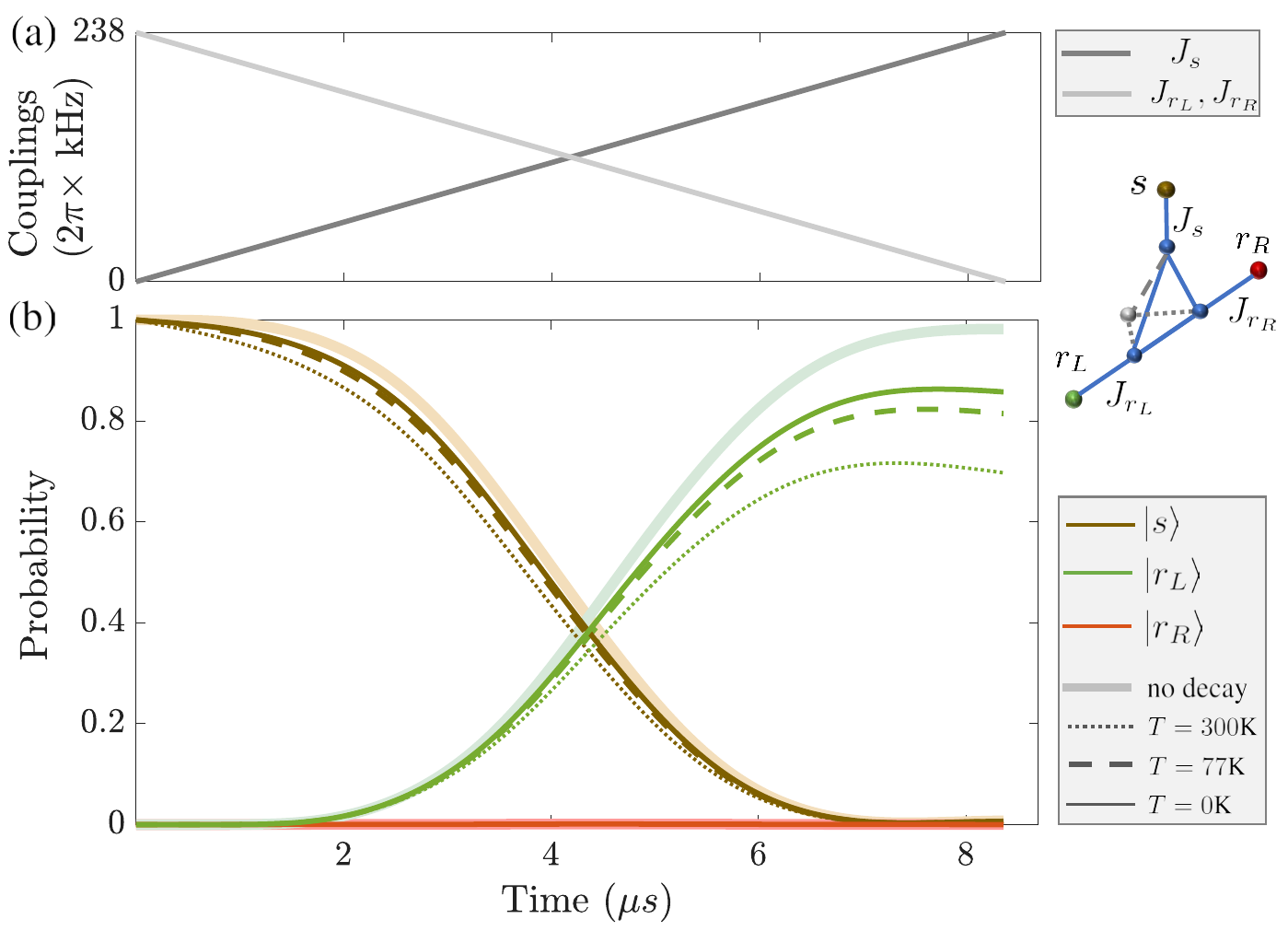}
\caption{(a) Temporal profile of the coupling strengths of the sender and receiver atoms coupled to the atoms of the flux triangle. The inset on the left illustrates the spatial configuration of the seven-atom router.
(b) Dynamics of excitation transfer from the sender $s$ to receiver $r_{L,R}$ atoms. The presence of auxiliary atom 4 results in nearly complete transfer to $r_{L}$ while the excitation probability of $r_{R}$ is less than $0.015$ at all times.  
The parameters of the flux triangle are the same as for the optimized solution in Fig.~\ref{fig:ryd} while the total decay rate of the atoms is $\Gamma_{\mathrm{tot}} \simeq 1/62,1/45,1/24\;\mu\mathrm{s}^{-1}$ for the temperatures $T=0,77,300\:$K (sold, dashed, dotted lines), respectively, with the lifetimes of the $nS$ and $nP$ states of Rb and Cs obtained from Ref. \cite{beterov2009quasiclassical}. 
For vanishing decay $\Gamma_{\mathrm{tot}}=0$ (thick faded line) the transfer is nearly perfect. }
\label{fig:decay}
\end{figure}

To demonstrate the operation of our scheme under realistic conditions, we have performed numerical simulations for the excitation transfer between the sender and receiver atoms attached to the flux triangle. The latter corresponds to the optimized configuration for four six-level atoms of Fig.~\ref{fig:ryd} in Sec.~\ref{sec:ed} with three main atoms 1,2,3 (Rb) and one auxiliary atom 4 or 5 (Cs). For simplicity, we consider only two states $\ket{0}$ and $\ket{1}$ of the sender and two receiver atoms (Rb) and hence the Hilbert space (of size $128+3$) is enlarged by three new states corresponding to the excitation on the sender or one of the two receiver atoms while all the atoms of the flux triangle are in the lower Rydberg states in the $n^{(\prime)}S$ manifold. The sender atom $s$ is coupled to atom 1 of the triangle with the time-dependent coupling $J_s(t)$ and the receiver atoms $r_{L,R}$ are coupled to atoms 2 and 3 with couplings $J_r(t)$, as illustrated in Fig.~\ref{fig:decay}.
The initial state is $\ket{\Psi(0)} = \ket{1_s} \otimes \ket{0_1,0_2,0_3} \otimes \ket{-_{4(5)}} \otimes \ket{0_{r_L},0_{r_R}}$ with only the sender atom excited to state $\ket{1}$ while all the other main atoms in state $\ket{0}$ and the auxiliary atom in state $\ket{-}$. We propagate the state vector of the system using the non-Hermitian Hamiltonian that includes the decay of the Rydberg states of the atoms outside the manifold of the computational Hilbert space. Since each atom in state $nS$ or $nP$ decays with the rate $\Gamma_{nS,nP}$ to other atomic states, the total decay rate is 
$\Gamma_{\mathrm{tot}} = \Gamma^{(\mathrm{Rb})}_{nP} + 5\Gamma^{(\mathrm{Rb})}_{nS} + \Gamma^{(\mathrm{Cs})}_{n'S} \simeq 6\Gamma^{(\mathrm{Rb})}_{nS} + \Gamma^{(\mathrm{Cs})}_{n'P}$. Then the norm of the wavefunction $\braket{\Psi(t)} \simeq e^{-\Gamma_{\mathrm{tot}}t}$ decreases with time reducing thereby the transfer probability. 

In Fig.~\ref{fig:decay} we show the results of our simulations. 
Neglecting the decay, we obtain nearly perfect transfer between the sender and the desired receiver atoms determined by the auxiliary atom 4. In the presence of Rydberg-state decay the rate of which depends on the temperature of the surrounding environment  
\cite{beterov2009quasiclassical}, the transfer selectivity is not affected but the final population of the desired state is reduced. We note that this population decrease of the total state determines the lower bound for the transfer probability, since in practice the decay of an atom which has already played its role in the transfer would  not decrease the population of the receiver atom. 

The seven-atom network represents the minimal setup for a quantum router. 
For larger networks of Rydberg atoms, other important considerations to attain good transfer fidelities include the influence and possible compensation of level shifts $\mu_i$ to attain resonance between the atoms in the flux triangle and in the chain, and taking into account that the resonant dipole-dipole interactions are long-range $\propto r^{-3}$ rather than nearest-neighbor, which affects the spectrum and dispersion of the chain (see Appendix \ref{sec:LSLR}).  

We finally note that the experimental parameters to realize our scheme are similar to those in recent experiments with Rydberg atoms \cite{lienhard2020realization,SSHRydberg2019}, but in principle we could go to Rydberg states with higher principal quantum numbers $n>70$ to further increase the coupling strengths and reduce the decay, which would improve the transfer fidelity for optimized parameters of the system.

\section{Summary} 
\label{sec:sum}

To conclude, we have presented an efficient method for coherent routing of quantum states in spin networks using quantum-state controlled chiral dynamics and non-dispersive propagation of spin excitations. 
Our routing scheme is based on quantum interference effects, whereby, depending on the state of the control qubit that controls the chirality of the router, a spin excitation is sent to the channel with 
constructive interference while destructive interference blocks the other channel. 
This should be contrasted with alternative schemes \cite{christensen2020coherent,qspintransistor2016,paganelli2013routing}
where controlled detuning of one or the other channel can facilitate the transfer of the excitation to the resonant channel and suppress its transfer to the non-resonant one. 
Depending on the particular physical setup of the spin or qubit network, one or the other methods may be more practical to implement. But we note that energy mismatch $\delta E >J$ can suppress the transfer probability only quadratically $(J/\delta E)^2$, while destructive interference can in principle be perfect, which is one of the potential advantages of the present and similar schemes.
We proposed a practical system involving laser-driven array of Rydberg atoms to implement this scheme.  
Our results can facilitate scalable quantum information processing and communication in large quantum registers of Rydberg atoms.

\acknowledgments 
We thank P. A. Kalozoumis and A. F. Tzortzakakis for useful discussions. N.E.P. and D.P. were supported by the EU QuantERA Project PACE-IN (GSRT grant No. T11EPA4-00015) and HORIZON-RIA Project EuRyQa (grant No. 101070144). 
D.P. was also supported by DFG through FOR-5413 QUSP.
S.O. and M.F. were supported by DFG through SFB-TRR 185 OSCAR (project No. 277625399).


\appendix 

\section{Dipole-dipole interaction} 
\label{sec:appDDI}

The static magnetic field $B$ along the $z$ direction defines the quantization axis and lifts the degeneracy of the magnetic sublevels of Rb and Cs atoms, $\delta E_{jm}=\mu_{B} B g_j m$, where the Lande factors are $g_{1/2} = 2$ and $g_{3/2} = 4/3$ for $nS_{1/2}$ and $nP_{3/2}$, respectively. 

The dipole-dipole interaction of Eq. (\ref{eq:dipole}) can be expanded as
\begin{equation} \label{eq:dd}
\begin{split}
V_{ij}= & \frac{1}{4\pi \epsilon_{0} |r_{ij}| ^{3}}\Big[ \hat{d}_{i}^{z} \hat{d}_{j}^{z}(1-3\cos^2{\theta_{ij}}) \\  
&+\frac{1}{2}( \hat{d}_{i}^{+} \hat{d}_{j}^{-}+ \hat{d}_{i}^{-} \hat{d}_{j}^{+})(1-3\cos^{2}{\theta_{ij}}) \\
&-\frac{3}{\sqrt{2}}( \hat{d}_{i}^{+} \hat{d}_{j}^{z}+ \hat{d}_{i}^{z} \hat{d}_{j}^{+})\sin{\theta_{ij}}\cos{\theta_{ij}}e^{-i\phi_{ij}} \\
&-\frac{3}{\sqrt{2}}( \hat{d}_{i}^{-} \hat{d}_{j}^{z}+ \hat{d}_{i}^{z} \hat{d}_{j}^{-})\sin{\theta_{ij}}\cos{\theta_{ij}}e^{i\phi_{ij}} \\  
& -\frac{3}{2}( \hat{d}_{i}^{+} \hat{d}_{j}^{+}e^{-2i\phi_{ij}}+ \hat{d}_{i}^{-} \hat{d}_{j}^{-}e^{2i\phi_{ij}})\sin^2{\theta_{ij}}\Big],
\end{split}
\end{equation}  
where $\hat{d}^{z,\pm}$ are dipole operators for the atomic transitions between states $\ket{nljm}$ and $\ket{n'l'j'm'}$
with $\Delta m = m' - m = 0,\pm 1$.
Then, in the truncated basis, the resonant exchange interactions between the main atoms $i$ and $j$, $\ket{0}_i\ket{1}_j \to \ket{1}_i\ket{0}_j$ with $\Delta m_i = +1$ and $\Delta m_j = -1$, and $\ket{1}_i\ket{0}_j \to \ket{0}_i\ket{1}_j$ with $\Delta m_i = -1$ and $\Delta m_j = +1$ are described by the second line of the above equation, while non-resonant exchange interactions between the main atom $i$ and auxiliary atom $j$, 
$\ket{1}_i \ket{-}_j \to \ket{0}_i\ket{+}_j$ with $\Delta m_i = -1$ and $\Delta m_j = -1$ and $\ket{0}_i \ket{+}_j \to \ket{1}_i\ket{-}_j$ with $\Delta m_i = +1$ and $\Delta m_j = +1$ are described by the last, fifth line of the above equation. 
When we simulate the complete system, we take into account all the dipole-allowed transitions with $\Delta m_{i,j} = 0, \pm 1$, leading to the Hilbert space of size 128 for a system of four six-level atoms sharing a single excitation. 

The matrix element of the dipole operator $\hat{\bm{d}}=e \hat{\bm{r}}$ for the transition between any two atomic states $\ket{nljm}$ and $\ket{n'l'j'm'}$,
\begin{equation}
e\bra{n^{\prime}l^{\prime}j^{\prime}m^{\prime}}\hat{\bm{r}}\ket{nljm}=
e \, C^{l^{\prime}j^{\prime}m^{\prime}}_{ljm} R^{n^{\prime}l^{\prime}}_{nl},
\end{equation}
is given by the product of the angular $C^{l^{\prime}j^{\prime} m^{\prime}}_{ljm}$ and radial $R^{n^{\prime}l^{\prime}}_{nl}$ parts \cite{sobelman2012atomic}. 
The angular part is calculated using the relation between the Clebsch-Gordan coefficients and the Wigner $3j$ and $6j$ symbols: 
\[
\begin{split}
C^{l^{\prime}j^{\prime}m^{\prime}}_{ljm}= & \sqrt{(2j^{\prime}+1)(2j+1)(2l^{\prime}+1)(2l+1)} \\
  & (-1)^{j-m+j^{\prime}+s +1}
  \begin{Bmatrix}
  j & 1 & j^{\prime} \\
  l^{\prime} & s & l
  \end{Bmatrix}
  \begin{pmatrix}
  l & 1 & l^{\prime} \\
  0 & 0 & 0
  \end{pmatrix}\\
  & \times
  \begin{pmatrix}
  j & 1 & j^{\prime} \\
  -m & -\Delta m & m^{\prime} 
  \end{pmatrix} ,
\end{split}
\]
where $s=1/2$ is the electron spin.  
The radial part 
\[
R^{n^{\prime}l^{\prime}}_{nl} =\int^{\infty}_{0} R_{nl}(r)\, R_{n^{\prime}l^{\prime}}(r) r^3 dr
\]
involves integration over radial wavefunctions $R_{nl}(r)$ and $R_{n'l'}(r)$ of the electron in the two states, and we employ the semi-classical approach \cite{d1991semiclassical} to calculate it for the transitions $nS_{1/2} \to nP_{3/2}$ between the Rydberg states of Rb and Cs.

\section{Router direction control}
\label{sec:RBGate}

\begin{figure}
\includegraphics[width=0.5\columnwidth]{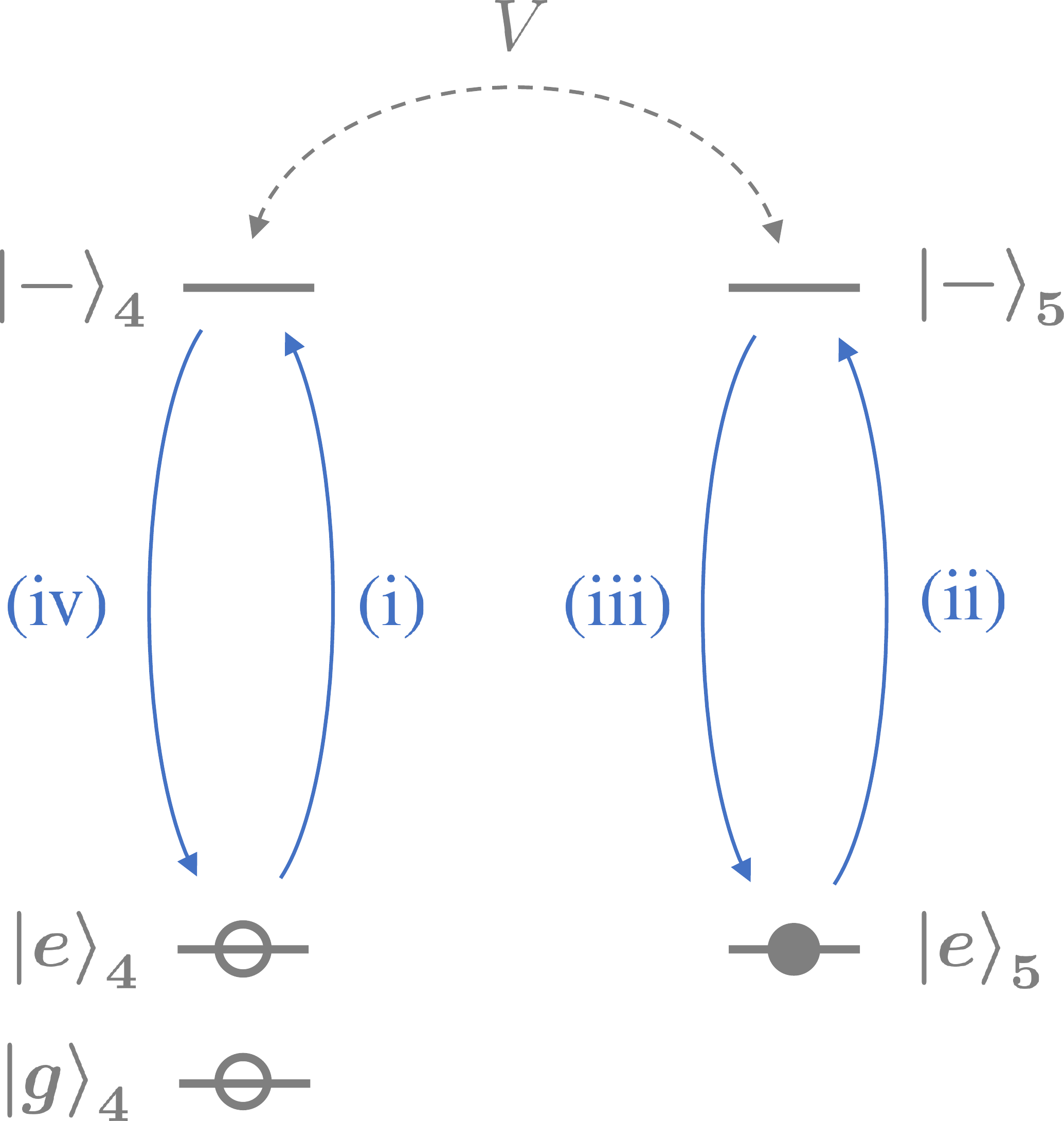}
\caption{Level scheme for the two auxiliary atoms 4 and 5 for realizing excitation routing by exciting one of the atoms to the Rydberg state $\ket{-}$ using an analog of the Rydberg blockade gate. 
The qubit that controls the routing direction is encoded in states $\ket{g}$ and $\ket{e}$ of atom 4.}
\label{fig:control}
\end{figure}

A control qubit determines which of the two auxiliary atoms 4 or 5 is prepared in the Rydberg state $\ket{-}$ to induce the corresponding chirality of $H_{\mathrm{eff}}$ with $\gamma_{\mathrm{tot}} = \mp \pi/2$.
The level scheme of the auxiliary atoms 4 and 5 is shown in Fig. \ref{fig:control}. The control qubit is encoded in the ground state sublevels $\ket{g}_4$ and $\ket{e}_4$ of atom 4, while atom 5 is prepared in state $\ket{e}_5$. To excite either atom 4 or 5 to the Rydberg state $\ket{-}$, we assume strong static dipole-dipole ($\nu=3$) or van der Waals ($\nu=6$) interactions $V=C_{\nu}/|r_{45}|^{\nu}$ between atoms 4 and 5 in the Rydberg state $\ket{-}$ and use a protocol similar to the standard Rydberg blockade gate with resonant laser pulses \cite{Lukin2001,SWMRMP2010}:
In step (i) we apply a $\pi$-pulse to atom 4 on the transition $\ket{e}_4 \to \ket{-}_4$: then atom 4 initially in state $\ket{g}_4$ or $\ket{e}_4$ (or any superposition thereof, $\ket{\psi}_4 = \alpha \ket{g}_4 + \beta \ket{e}_4$) ends up in state $\ket{g}_4$ or $i\ket{-}_4$ (or any superposition thereof, $\ket{\psi}_4 = \alpha \ket{g}_4 + i\beta \ket{-}_4$). In step (ii) we apply a $\pi$-pulse with Rabi frequency $\Omega \ll V$ to atom 5 on the transition $\ket{e}_5 \to \ket{-}_5$: then atom 5 is transferred to state $i\ket{-}_5$ if atom 4 was in state $\ket{g}_4$, otherwise atom 5 remains in state $\ket{e}_5$ since its transition $\ket{e}_5 \to \ket{-}_5$ is suppressed by the strong interaction $V$ between the Rydberg levels $\ket{-}$. Hence, only one of the two atoms, 4 or 5, is prepared in state $\ket{-}$, or their single-excitation superposition, 
\[
\ket{\psi}_{45} = i \alpha \ket{g}_4 \ket{-}_5 + i \beta \ket{-}_4 \ket{e}_5.
\]
We note that while the two auxiliary atoms should be 
sufficiently close to each other for the blockade to be effective, they should also be sufficiently far apart so that each of them could be separately addressed by a laser. 
We also note that we neglect the diagonal (static dipole-dipole or van der Waals) interactions between the auxiliary atom and the main atoms of the chain. 

We can now perform the excitation routing protocol as described in the main text. After the routing is complete, we perform steps (iii) and (iv) by applying $\pi$-pulses to atoms 5 and 4 to transfer them back to the initial state. 
More precisely, the initial and final states of the two auxiliary atoms and the sender and receiver spins are 
\[
\begin{split}
(\alpha \ket{g}_4 + \beta \ket{e}_4) \otimes \ket{e}_5 \otimes \ket{1}_s \otimes \ket{0}_{r_L} \otimes \ket{0}_{r_R}  \\
\to i (\alpha \ket{g}_4 \ket{-}_5 + \beta \ket{-}_4 \ket{e}_5) \otimes \ket{1}_s \otimes \ket{0}_{r_L} \otimes \ket{0}_{r_R}
\\
\to - (\alpha \ket{g}_4 \otimes \ket{0}_{r_L} \otimes \ket{1}_{r_R} + \beta \ket{e}_4 \otimes \ket{1}_{r_L} \otimes \ket{0}_{r_R}) \\
\otimes \ket{e}_5\otimes \ket{0}_{s},
\end{split}
\]
i.e., we obtain, in general ($\alpha \beta \neq 0$), an entangled state of the control qubit and the two receiver qubits.

\section{Time-dependent couplings $J_{s,r}$}
\label{sec:Jtsr}

\begin{figure}
\includegraphics[width=0.7\columnwidth]{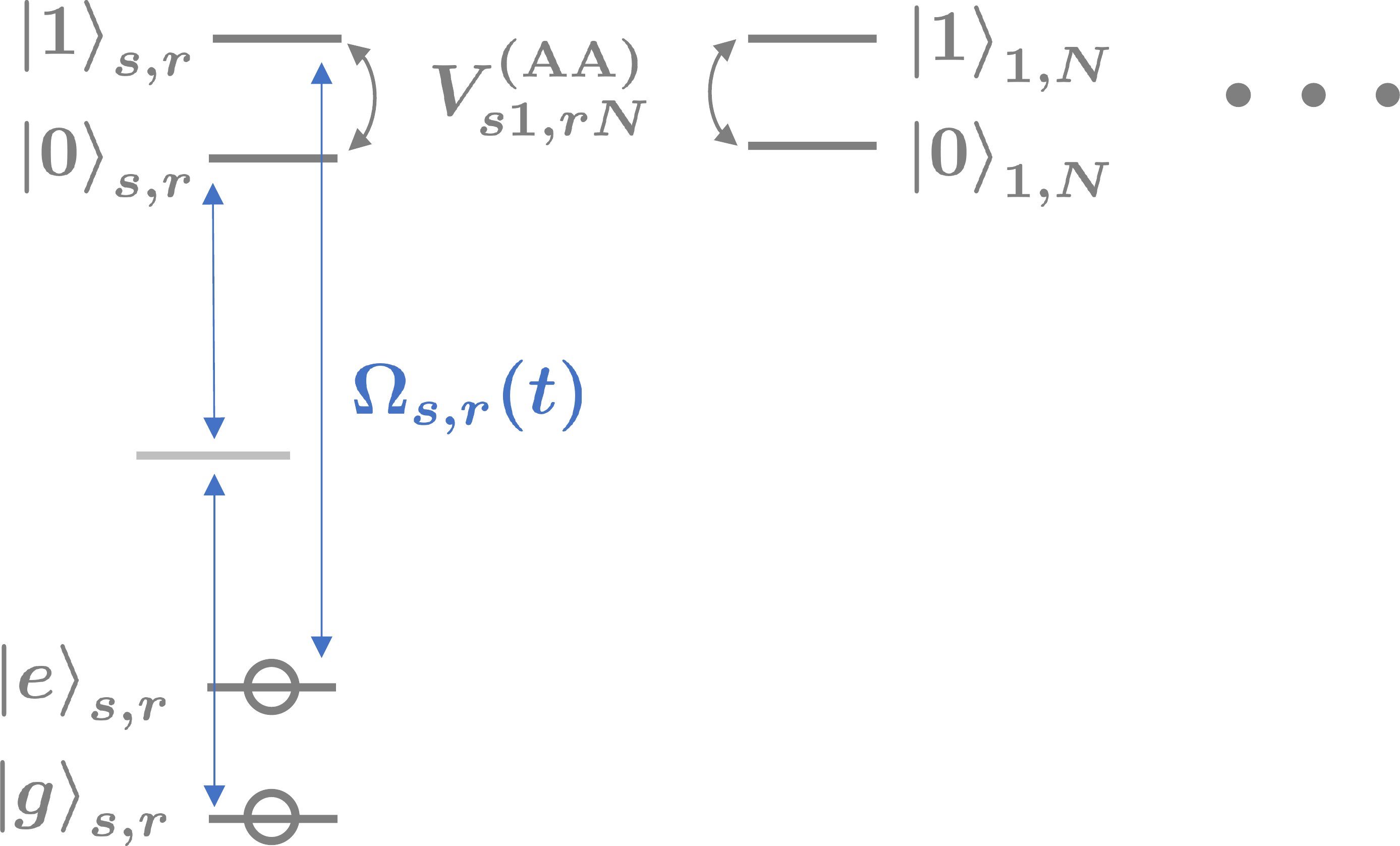}
\caption{Level scheme of the sender and receiver spins to implement their time-dependent couplings to the spin network of Rydberg atoms.}
\label{fig:Jsr}
\end{figure}

We assume cold atoms trapped in an array of microtraps. 
The qubit state to be transferred is encoded in the ground state superposition of the sender atom, $\ket{\psi}_s = c_0 \ket{g}_s + c_1 \ket{e}_s$, while the receiver atom(s) and the rest of the atoms in the network are in state $\ket{g}$. We first apply resonant $\pi$-pulses to the sender, receiver(s) and the network atoms on the transition $\ket{g} \to \ket{0}$ to prepare them in the lower Rydberg state $\ket{0}$. Since $\ket{0}$ is an $nS_{1/2}$ Rydberg state, its excitation from the ground state sublevel $\ket{g}$ requires a two-photon transition via an intermediate $5P_{3/2}$ or $6P_{3/2}$ state of Rb. 
Next, we apply to the sender atom a (near-)resonant laser pulse $\Omega_s(t)$ on the transition $\ket{e}_s \to \ket{1}_s$ to the excited Rydberg state, see Fig.~\ref{fig:Jsr}. This is a one-photon UV transition $5S_{1/2} \to nP_{3/2}$. Since the Rydberg transition $\ket{1}_s \to \ket{0}_s$ is coupled via the exchange interaction with transition $\ket{0}_1 \to \ket{1}_1$ of the first atom of the chain, the time-dependence of the coupling laser pulse determines the time-dependence of the effective coupling $J_s(t) \propto \Omega_s(t)$, as in stimulated Raman 
processes \cite{STIRAP-RMP1998,vitanov2017stimulated}.
The single-excitation wavepacket will then propagate in the spin network of the Rydberg atoms, until it reaches the receiver atom initially in state $\ket{0}_r$, which is then transferred to the qubit state $\ket{e}_r$ by another laser pulse $\Omega_r(t)$ acting on the transition  $\ket{1}_r \to \ket{e}_r$ in the time-reversed manner. Finally, we apply resonant $\pi$-pulses on the transitions $\ket{0} \to \ket{g}$
of the sender, receiver and the network atoms to bring them to the trapped ground electronic states.

\section{Level shifts and interaction range of Rydberg atoms}
\label{sec:LSLR}

\begin{figure}[t]
\includegraphics[width=0.9\columnwidth]{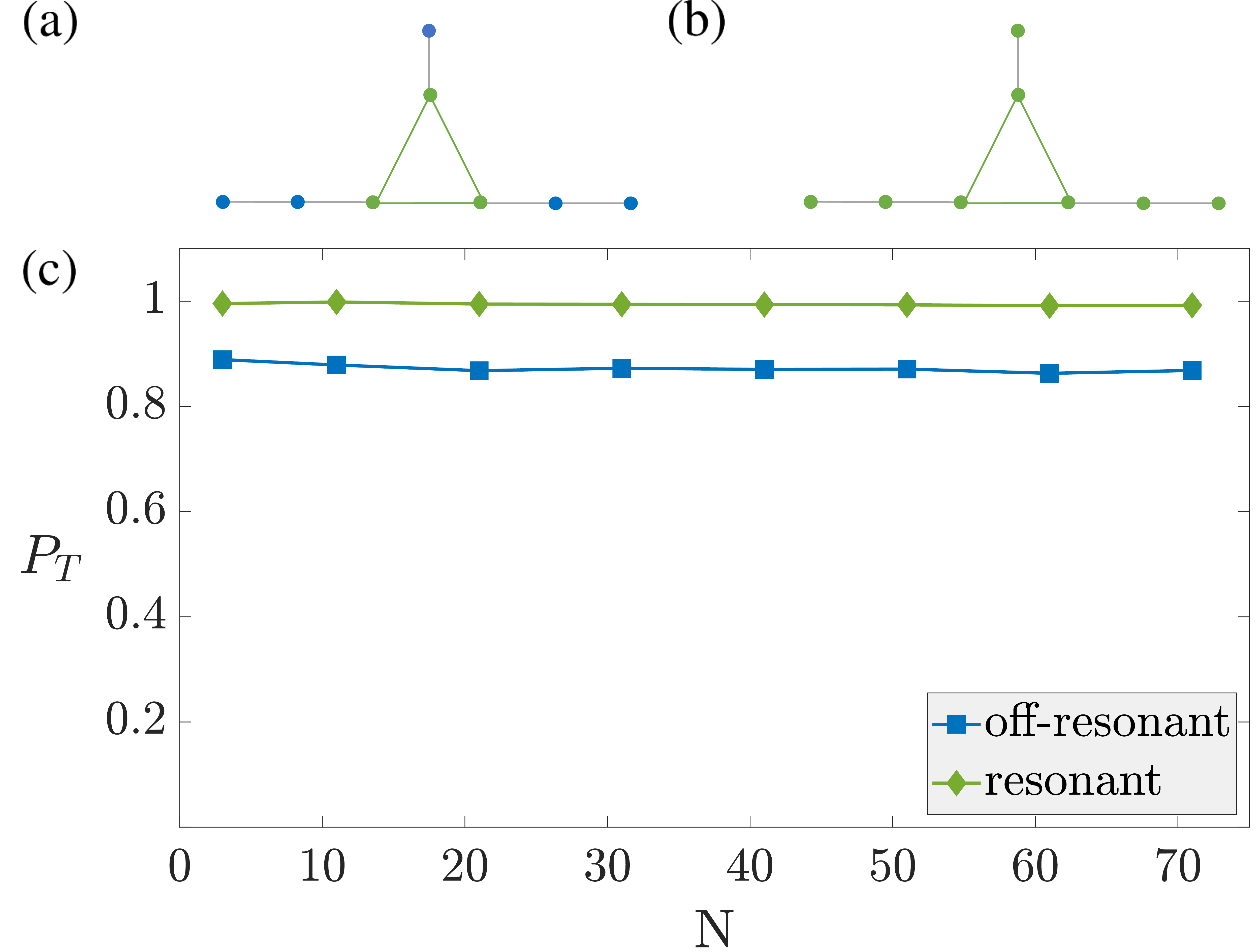}
\caption{Schematics of the router setup with (a) the atoms of the flux triangle slightly off-resonant with respect to the atoms of the chain, and (b) all the atoms of the triangle and chain are resonant. 
(c) The corresponding transfer probabilities $P_T$ between the sender and receiver spin via the chain of $N$ spins.}
\label{fig:comp}
\end{figure}

The second-order level shifts $\mu_{1,2,3}$ of the atoms in the flux triangle make them slightly off-resonant with the network as well as the sender and receiver atoms. This reduces the fidelity of excitation transfer between the sender and receiver atoms, as shown in Fig.~\ref{fig:comp}(a,c). 
But if we compensate these levels shifts using, e.g., non-resonant lasers to induce ac Stark shifts of the atomic levels, we recover the nearly perfect state transfer in the network of Rydberg atoms, see Fig.~\ref{fig:comp}(b,c)

\begin{figure}[t]
\includegraphics[width=0.9\columnwidth]{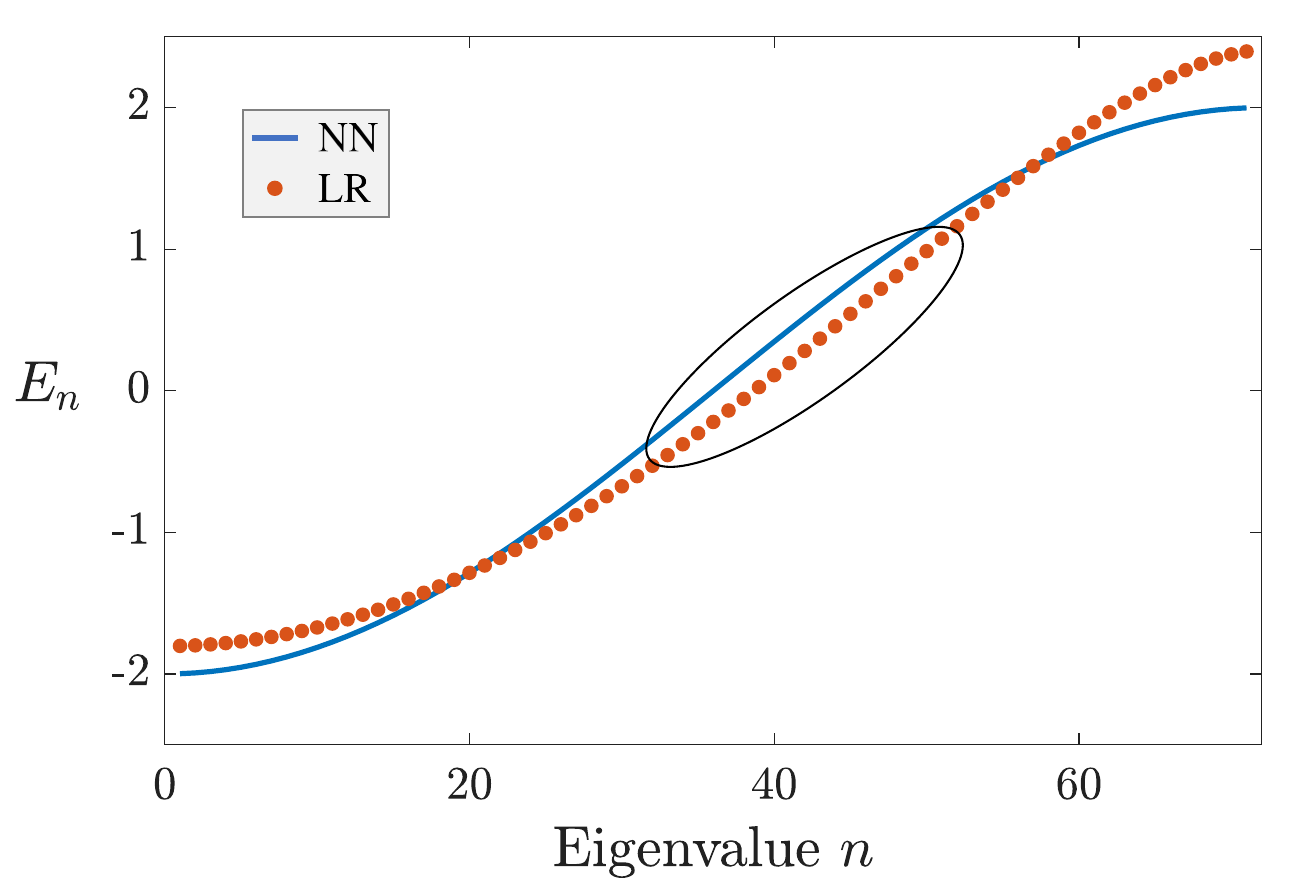}
\caption{Single-excitation spectrum $E_n$ (is in units of $J$) of a chain of $N=71$ spins with nearest-neighbor (NN) exchange interactions (blue solid line) and longer-range (LR) $1/r^3$ interactions (red dotted line).}
\label{fig:spec}
\end{figure}

In Fig.~\ref{fig:comp} we also assume that the distance between the neighboring atoms in the network is chosen such that their (real) exchange coupling $J$ is equal to the absolute value $\abs{J_{12,13,23}}$ of the effective couplings between the atoms of the flux triangle. But since the exchange couplings are due to the resonant dipole-dipole interactions scaling with the interatomic distance $r$ as $C_3/r^3$, the single excitation spectrum of the chain deviates from the simple cosine law of Eq.~(\ref{eq:cos}) as \cite{Piil2007,Letscher2018}
\begin{equation}
E_{n} = 2 \sum_{m=1}^{N-1} \frac{J}{m^3}\cos{\frac{\pi n m}{N+1}}, \label{eq:sumcos}
\end{equation}
where the $m=1$ term corresponds to the nearest-neighbor exchange, while terms with $m=2,3,\ldots$ stem from the longer-range exchange interactions between the next-nearest-neighbors, next-next-nearest-neighbors, etc. The linear part of the spectrum of the chain is then displaced towards higher energies, see Fig. \ref{fig:spec}. Hence, to launch (and absorb) a non-dispersive wavepacket, the energies (effective magnetic fields $B_{s,r}$) of the sender and receiver spins should be accordingly tuned.

\end{document}